\documentclass[prl,aps,showpacs,floatfix,twocolumn, superscriptaddress]{revtex4-1}
\usepackage{color}
\usepackage{colordvi}
\usepackage{graphicx,amssymb,epsfig,amsmath}
\usepackage{epstopdf}
\usepackage{amsmath}
\usepackage{bm}
\usepackage{amsfonts}
\usepackage[bottom]{footmisc}
\begin{document}

\title{Detecting One-Dimensional Dipolar Bosonic Crystal Orders via Full Distribution Functions }
\author{Budhaditya Chatterjee}
\email{bchat@iitk.ac.in}
\affiliation{Department of Physics, Indian Institute of Technology-Kanpur, Kanpur 208016, India}
\author{Camille L\'ev\^eque}
\email{camille.leveque@tuwien.ac.at}
\affiliation{Vienna Center for Quantum Science and Technology, Atominstitut, TU Wien, Stadionallee 2, 1020 Vienna, Austria}
\affiliation{Wolfgang Pauli Institute c/o Faculty of Mathematics, University of Vienna, Oskar-Morgenstern Platz 1, 1090 Vienna, Austria}
\author{J\"org Schmiedmayer}
\email{Schmiedmayer@atomchip.org}
\affiliation{Vienna Center for Quantum Science and Technology, Atominstitut, TU Wien, Stadionallee 2, 1020 Vienna, Austria}
\author{Axel U. J. Lode}
\email{auj.lode@gmail.com}
\affiliation{Institute of Physics, Albert-Ludwig University of Freiburg, Hermann-Herder-Strasse 3, 79104 Freiburg, Germany}
\affiliation{Vienna Center for Quantum Science and Technology, Atominstitut, TU Wien, Stadionallee 2, 1020 Vienna, Austria}

\begin{abstract}

We explore the groundstates of a few dipolar bosons in optical lattices with incommensurate filling. The competition of kinetic, potential, and interaction energies leads to the emergence of a variety of crystal state orders with characteristic one- and two-body densities. We probe the transitions between these orders and construct the emergent state diagram as a function of the dipolar interaction strength and the lattice depth. We show that the crystal state orders can be observed using the full distribution functions of the particle number extracted from simulated single-shot images. 
\end{abstract}

\maketitle

The realization of Bose-Einstein condensates (BECs) of dipolar atoms~\cite{griesmaier05,beaufils08,lu11,aikawa12,ni08,zwierlein15} and molecules~\cite{ni08,zwierlein15} provides new perspectives to study phase transitions in correlated quantum systems~\cite{baranov08,lahaye09}. The anisotropic and long-ranged dipole-dipole interactions lead to a plethora of new phenomena absent in conventional BEC, e.g., directional elongation~\cite{yi01,santos00,goral00} and geometric stabilization~\cite{yi01,santos00,eberlein05,goral02a,koch08}. A lower dimensionality results in additional physical features: $p$-wave superfluidity~\cite{bruun08, cooper09}, Luttinger-liquid-like behavior~\cite{arkhipov05,citro07,depalo08,pedri08}, and anisotropy in curved geometries~\cite{zollner11,zollner11pra,maik11}.

Atoms in optical lattices are experimentally very tunable and serve as quantum simulators for condensed matter systems~\cite{bruder,greiner02,Liberto16,Kock16,Kock15,landig15,jotzu14,Landini18,Langen15,Schweigler17,Engelsen17,lahaye10,anna13,fischer13,gallemi13,gallemi16}. Few-particle systems especially provide an experimental bottom-up access to many-body physics~\cite{selim1,selim2,selim3}. The interplay between anisotropic long-range and contact interactions results in the emergence of new phases beyond the usual superfluid ($SF$) and Mott-Insulator ($MI$). A density-wave phase ($DW$)~\cite{goral02,dalla06,biedron18,Maluckov12}, characterized by an alternate filling of lattice sites and a supersolid phase~\cite{Kovrizhin2005,Yi2007,Grimmer2014,Cinti2016,Danshita2009,biedron18}, with coexistent $DW$ and superfluidity were observed. Exotic phases such as Haldane insulators~\cite{dalla06, deng11}, checkerboards~\cite{goral02,menotti07} and Mott solids~\cite{zoller10} were predicted. Remarkably, a crystal state ($CS$) emerges for dominant dipolar interactions~\cite{arkhipov05,astrakharchik08b,deuretzbacher10,Schachenmayer10,bera18,zollner11,zollner11pra,chatterjee12,chatterjee17a,chatterjee17b}.

In this Letter, we establish protocols to detect the remarkable plethora of crystal orders emerging in lattices incommensurately filled with dipolar bosons. 
The physics of the crystal state cannot be addressed using the Hubbard model~\cite{Note_Hubbard} and requires an ab-initio many-body description~\cite{chatterjee17a,chatterjee17b}.
We study groundstates of a few dipolar atoms in one-dimensional lattices by numerically solving the full Schr\"odinger equation with the multiconfigurational time-dependent Hartree method for bosons~\cite{alon08,streltsov07,ultracold,axel1,axel2,lode:19}. 

As Refs.~\cite{chatterjee15,cao17,Mistakidis17,Koutentakis17,Mistakidis18a,Streltsov13,Streltsova14,Klaiman18,Marchukov19}, we explore the full range of interaction strengths to observe emergent distinct crystal orders, due to the competition between kinetic, potential, and interaction energies.
Our main finding is that all these crystal orderings and thereby the phase diagram can be 
unequivocally characterized using the full distribution functions of the position-dependent particle number operator extracted from simulated experimental absorption or single-shot images~\cite{sakmann16,lode17,tsatsos17,chatterjee17a}.  We emphasize that our results are experimentally feasible as single-shot images with single-atom sensitivity for few-atom systems has been observed \cite{Bakr:09,Sherson:10}.
With our present finite-size considerations, we cannot make claims about the true quantum phases in the thermodynamic limit. However, we demonstrate that the crystal orders are valid for the state diagram, the finite-size analog of the thermodynamic phase diagram for different lattice sizes, particle numbers, and different types of boundary conditions~\cite{SI}.

Consider polarized, dipolar bosons in a quasi-one-dimensional lattice potential having tight transversal confinement of characteristic length $a_\perp$. 
The $N$-body Hamiltonian reads
\begin{equation}
H=\sum_{i=1}^{N} \left[ T_i + V_{ol}(x_i) \right] + \sum_{i<j} V_{int}(x_i-x_j).
\end{equation} 
Here $T_i$ is the kinetic energy of the $i$-th boson, $V_{ol}= V\sin^{2}(\kappa x)$ is the lattice potential with a depth $V$ and a wave-vector $\kappa$. Hard wall boundaries at $x=\pm S\pi/2\kappa$ restrict the lattice to $S$ sites. $V_{int}(x_i-x_j) = \frac{g_d}{\vert x_i - x_j\vert^3 + \alpha}$ is the (purely) dipolar interaction of strength $g_d$ ~\cite{Note1}. The transversal confinement introduces the short-scale cutoff $\alpha \approx {a_\perp}^3$~\cite{sinha07,deuretzbacher10,cai10,Note2}.  
 All quantities are given in terms of recoil energy \cite{Note_recoil}. We consider a cutoff $\alpha=0.05$, $N=8$ bosons, and $S=5$ lattice sites~\cite{Note_parameter}.
 
\begin{figure}
	\centering
	\includegraphics[width=1.0\linewidth]{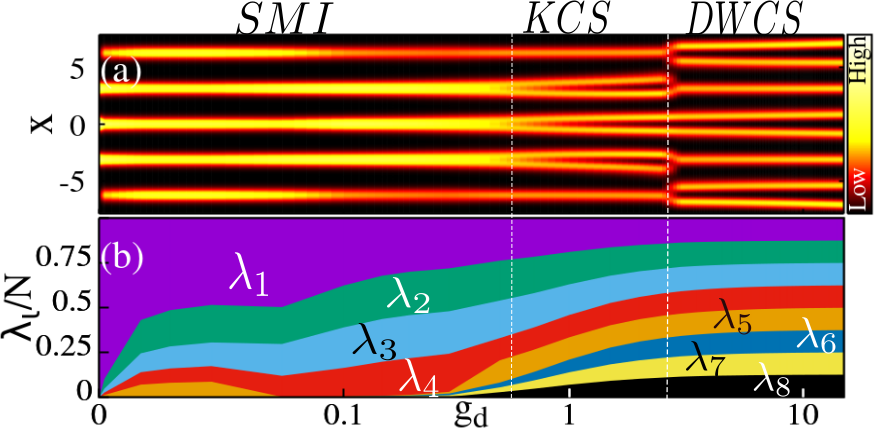}
	\caption {\textbf{(a) One-body density $\rho(x)$ as a function of the dipolar interaction strength $g_d$ for a lattice depth $V=8$.} For the $SMI$, $g_d\lesssim 0.8$, the density exhibits a five-fold structure. 
		As $g_d$ increases, the density develops a twofold splitting in the doubly-occupied central three wells displaying the onset of $SMI \rightarrow KCS$ transition. 
		For even larger $g_d$, the density transitions to a pattern with alternating single and double occupations signifying the $DWCS$ transition. See SI, Sec.~S2~\cite{SI} for $\rho(k)$ in momentum space.
		\textbf{(b) Natural occupations (plotted cumulatively) as a function of $g_d$.} For small $g_d$, the $SF$ fraction results in the dominance of $\lambda_1$. With increasing $g_d$, the fragmentation increases as several $\lambda_{k>1}$ become significant. Beyond the $SMI \rightarrow KCS$ transition, the system is maximally fragmented with $8$ orbitals populated almost equally.}
	\label{fig:den1d}
 \end{figure}

We now discuss possible crystal-state orderings using the density $\rho(x)=\langle \Psi \vert \hat{\Psi}^\dagger(x) \hat{\Psi}(x) \vert \Psi \rangle$ as a function of the interaction strength $g_d$ [Fig. \ref{fig:den1d}(a)]. For $g_d=0$, we obtain a pure $SF$. The incommensurate setup implies the absence of a pure $MI$; as $g_d$ increases we find an $MI$ which coexists with an $SF$ ($SMI$)~\cite{brouzos10}. Observed from $\rho(x)$: the two outer wells have a smaller population than the central wells; $N=5$ atoms in the $MI$ coexist with $N=3$ bosons in the $SF$ fraction. The $SF$ localizes in the central wells to minimize the kinetic energy with hard-wall boundary conditions.

The crystal transition occurs at $g_d \approx 0.8$: due to their repulsive dipolar interactions, the bosons avoid each other and save interaction energy minimizing their overlap.
The density [Fig.~\ref{fig:den1d}(a)] splits in the doubly occupied central wells -- a signature of the crystal transition. The crystal state's site occupations are $1,[11],[11],[11],1$, where $[11]$ denotes the two-hump density in doubly-occupied lattice sites. Since the double occupation of the three central wells results from the kinetic energy in the Hamiltonian, we term this state ``kinetic crystal state'' ($KCS$). 

A further increase of $g_d$ makes the interaction energy overcome the kinetic energy; the double occupation of adjacent sites is energetically unfavorable [Fig. \ref{fig:den1d}(a)]: instead of nearest neighbors, next-nearest neighbors are now doubly occupied in a density-wave-ordered structure at $g_d\gtrsim 3$.
The ``standard'' density-wave-order for our system would have the occupations $2,1,2,1,2$ (Ref.~\cite{Maluckov12}). Due to the strong dipolar interactions, the bosons' density in doubly occupied sites is spatially split and a density-wave crystal state ($DWCS$) with occupations $[11],1,[11],1,[11]$ results. The $DWCS$ represents a completely new crystal order, featuring the coexistence of $DW$ arrangement and crystallization. The $KCS$ and the $DWCS$ are possible crystal orders and hence a subset of a general crystal state $CS$. Here, and henceforth, we use the label $CS$ for a crystal state in either the $KCS$ or the $DWCS$ arrangement. 
See Ref.~\cite{SI}, Secs. S4, S5, and S9, for finite-size and boundary-condition effects on the crystal orders; importantly, the $KCS$ fades away for a larger system size while the $DWCS$ and $CS$ prevail.

Various crystal arrangements of the atoms with respect to the lattice can be obtained from a purely classical model (Sec.~S4~\cite{SI}). However, many-body calculations are necessary to capture the quantum properties of the crystal states: many modes contribute to the quantum field of the crystal states [Fig.~\ref{fig:den1d}(b)] -- their properties are only accessible via a realistic model of the many-body wavefunction. Moreover, the $SF$ and $MI$ are purely quantum states and the transition from such a non-crystal to a crystal order cannot be obtained from a classical model.

\begin{figure}
	\centering
	\includegraphics[width=1.0\linewidth]{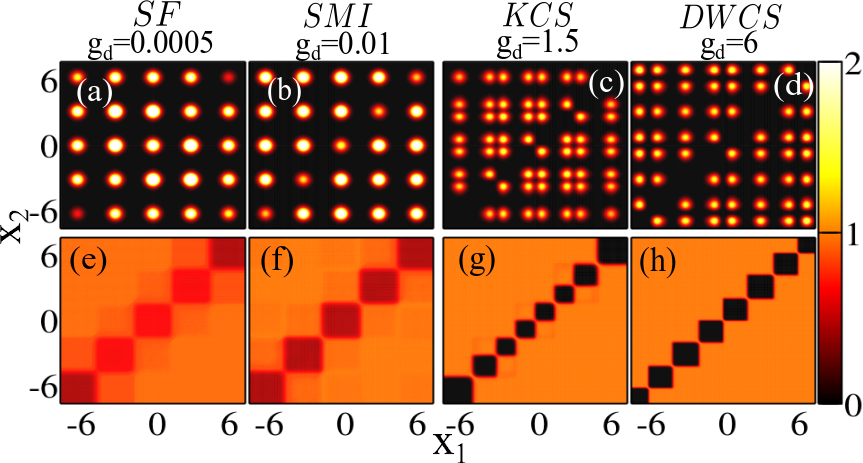}
	\caption { \textbf{(a)--(d) Two-body density $\rho^{(2)}(x_1,x_2)$.} (a) $g_d = 0.0005$: The pure $SF$ shows a nearly square-lattice-like equispaced distribution. (b) $g_d = 0.01$: The localization of bosons in the $SMI$ results in a diagonal depletion. (c) $g_d = 1.5$: For $KCS$, a correlation hole develops; the probability to find two bosons at the same position vanishes. (d) $g_d = 6.0$: The $DWCS$ shows a completely split but non-uniform density with density-wave order.
	\textbf{(e)--(h): $2^{nd}$ order spatial correlation function [$|g^{(2)} (x_1,x_2)|$]}.(e) $g_d = 0.0005$: the bosons are significantly coherent for $SF$. (f) $g_d = 0.01$: reduction of off-diagonal coherence for $SMI$. (g) $g_d = 1.5$: diagonal coherent blocks splits as the bosons crystalize at $KCS$.  (h) $g_d = 6$ Completely split coherent blocks centered at each boson position. 
		}
	\label{fig:den2d}
\end{figure}

To explain the emergent many-body properties and the localization in the observed states, we discuss the two-body density $\rho^{(2)}(x_1,x_2)= \langle \Psi \vert \hat{\Psi}^{\dagger}(x_1)\hat{\Psi}^{\dagger}(x_2) \hat{\Psi}(x_1) \hat{\Psi}(x_2) \vert \Psi \rangle$ for characteristic values of $g_{d}$ [Fig.~\ref{fig:den2d}(a--d)]. Unlike the particle arrangement corresponding [one-body density, Fig.~\ref{fig:den1d}(a)], the analysis of the two-body density exhibits the many-bodyness of the crystal state: the two-body densities in Fig.~\ref{fig:den2d}(b)--(d)  -- unlike for classical, semi-classical, and mean-field models -- is not a product of one-body densities.

For small interaction strength ($g_d = 0.0005$) in the $SF$ the maxima of $\rho^{(2)}$ are nearly uniformly distributed at positions $(x_1,x_2)$ in the vicinity of the minima of the lattice [Fig.~\ref{fig:den2d}(a)]. For larger interaction, $g_d = 0.01 $, a partial depletion along the diagonal ($x_{1},x_2 \approx x_{1}$) of $\rho^{(2)}$ occurs. This depletion results from the formation of an $MI$ that coexists with an $SF$ [Fig.~\ref{fig:den2d}(b)]. At stronger interactions, $g_d = 1.5$, the bosons in the doubly-occupied central wells crystallize forming the $KCS$. The diagonal of $\rho^{(2)}(x_{1},x_{1})\approx0$ is completely depleted: a correlation hole is formed, the probability of detecting two bosons at the same position vanishes. The split maxima in the three central wells result from the on-site interaction-driven splitting $2 \rightarrow [11]$.
For $g_{d}=6$, in the $DWCS$ [Fig.~\ref{fig:den2d}(d)], a split inter-site structure of the two-body density $\rho^{(2)}$ is present for every odd site of the lattice potential. The diagonal depletion is wider compared to the $KCS$ [Fig.~\ref{fig:den2d}(c),(d) for $x_1\approx x_2$]. The non-uniform distribution of the maxima of $\rho^{(2)}$ heralds the $DWCS$.

For extremely strong interactions $g_d > 50$ (not shown), the $DWCS$ transforms into a crystal state $CS$. To minimize the interaction that overwhelms the potential the distribution of the maxima of both $\rho(x)$ and $\rho^{(2)}(x_1,x_2)$ becomes equidistant [~\cite{SI}, Sec.~S3]. This $CS$ differs from the $SMI$, $KCS$, and the $DWCS$, because the distribution of maxima of the densities is \textit{not} dictated by the minima of the lattice potential (~\cite{SI},Sec.~S4,S5,S9) for other possible crystal orders at different particle numbers, lattice sizes and boundary conditions.

To explore the coherence properties of the emergent states, we analyze the $2^{nd}$ order spatial (Glauber) correlation functions~\cite{sakmann08,Glauber63}, defined as $g^{(2)}(x_1,x_2,x_{1}^{\prime},x_{2}^{\prime}) = \frac{ \rho^{(2)}(x_{1}, x_{2},x_{1}^{\prime}, x_{2}^{\prime}) }{\sqrt{\rho(x_{1})\rho(x_{1}^{\prime})\rho(x_{2})\rho(x_{2}^{\prime}) }}$ and quantifies the $2^{nd}$ order coherence in the system. Fig.~\ref{fig:den2d}(e)--(h) display the diagonal $|g^{(2)}|\equiv|g^{(2)}(x_1,x_2)| = |g^{(2)}(x_1,x_2,x_1^\prime=x_1,x_2^\prime=x_2)|$. 

In the $SF$ ($g_d = 0.0005$) the system is delocalized and shows substantial coherence [$|g^{(2)}|\approx1$ almost throughout Fig.~\ref{fig:den2d}(e)]. For the $SMI$, $g_d = 0.01$, a partial localization in the lattice sites reduces the off-diagonal coherence, reflected in the diagonal anticorrelation blocks $|g^{(2)}|< 1$, Fig.~\ref{fig:den2d}(f). For $g_d =1.5$ ($KCS$), the diagonal coherent blocks split as the bosons crystallize. The (anti) correlation regions are now centered on each boson, [Fig.~\ref{fig:den2d}(g)]; this persists also for the $DWCS$ [$g_d =6$, Fig.~\ref{fig:den2d}(h)], showing the complete localization and decoherence of atoms in crystal states. These coherence properties are inherently a quantum many-body effect absent in mean-field approaches (~\cite{SI}, Sec.~S8). The two-body correlation functions and densities in Fig.~\ref{fig:den2d}, as expected for multiple significant eigenvalues of the reduced one-body density matrix [Fig.~\ref{fig:den1d}(b) and below] demonstrate the appealing many-body structure of crystallized bosons.

\begin{figure*} 
	\centering 
	\includegraphics[width=2\columnwidth]{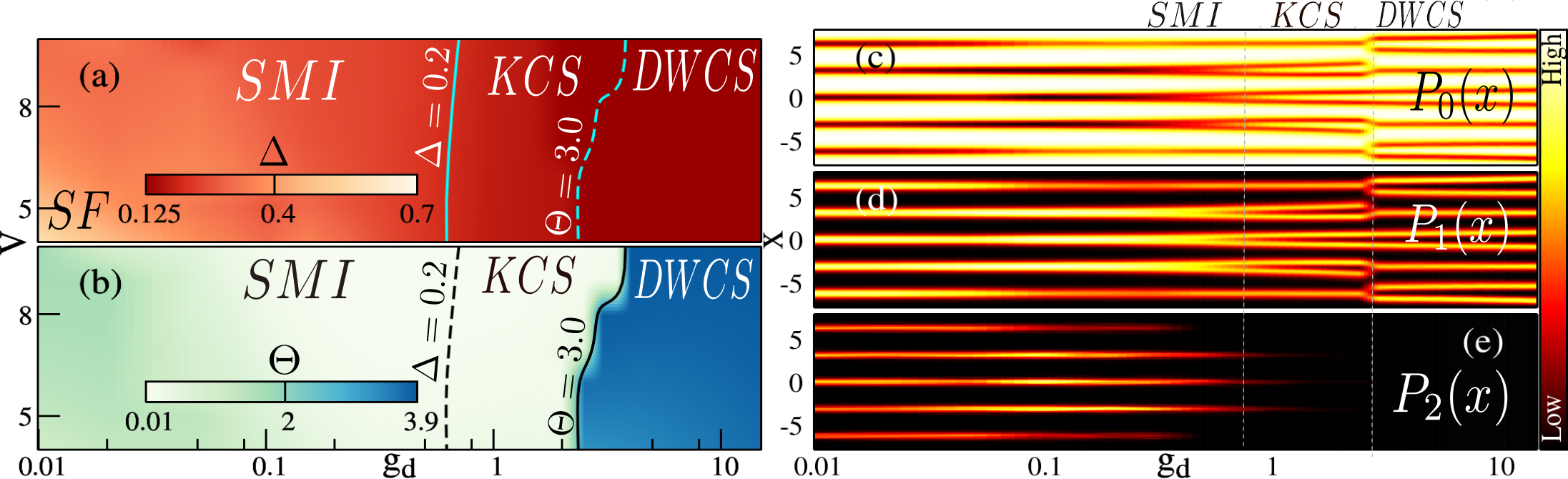}
	\caption {\textbf{Characterization and detection of the $KCS$ and $DWCS$.} (a) \textbf{Crystal order parameter} $\Delta$ as a function of the interaction strength $g_d$ and lattice depth $V$. The maximum, $\Delta = 1$, corresponds to the condensed $SF$. The $SMI$ is revealed by intermediate $\Delta<1$. The minimum value, $\Delta=0.125$, corresponds to a $CS$ thereby characterizing its formation. (b) \textbf{Imbalance parameter} $\Theta$ as a function of $g_d$ and $V$. Both $SMI$ and $KCS$ correspond to low values of $\Theta$, while the maximum value indicates the $KCS \rightarrow DWCS$ transition. 
	(c--e) \textbf{Full distribution functions $P_n(x)$} as a function of $g_d$ for $V=8$ evaluated from $10,000$ single-shots. The plots of the probability to detect zero [$P_0(x)$] or one [$ P_{1}(x)$] particle are reminiscent of the one-body density [compare Fig.~\ref{fig:den1d}(a)]. The probability of detecting two particles [$P_2(x)$] vanishes clearly when the $CS$ transition occurs. At large interactions, $g_d\gtrsim 6$, $P_0(x)$ and $P_1(x)$ exhibit a density-wave pattern. This pattern along with vanishing $P_2(x)$ reveals the $DWCS$. See SI,~\cite{SI} Sec.~S6, for full distribution functions in momentum space.}
	\label{fig:phase_fdf} 
\end{figure*} 

The $SMI\rightarrow KCS$ and $KCS \rightarrow DWCS$ transitions depend on $g_d$ and $V$; we construct the state diagram for these parameters. The $SMI \rightarrow CS$ transition can be determined from a crystal state order parameter $\Delta$~\cite{chatterjee17a}, 
\begin{equation}
\Delta= \sum_k \left(\frac{\lambda_k}{N}\right)^2. \label{OP}
\end{equation}
The $\lambda_k$ are the $k^{th}$ eigenvalues (natural occupations) obtained from diagonalizing the one-body reduced density matrix,
\begin{equation}
\rho^{(1)}(x,x')=\langle \Psi \vert \hat{\Psi}^{\dagger}(x) \hat{\Psi}(x') \vert \Psi \rangle=\sum_i \lambda_i \varphi^*_i(x) \varphi_i(x^{\prime}), \label{RDM}
\end{equation}
together with the eigenfunctions $\varphi_i(x)$ (natural orbitals). 

The values $\lambda_k$ determine if the system is condensed: one natural occupation is macroscopic ($\lambda_1 \approx N$) ~\cite{penrose56} or fragmented: multiple $\lambda_k$'s are macroscopic~\cite{spekkens,mueller}. Fig.~\ref{fig:den1d}(b) shows $\lambda_k$ as a function of $g_d$. For $0<g_d<0.002$, the system is condensed $SF$ and $\lambda_1 \approx N$. For $0.002<g_d<0.8$, the system fragments and multiple $\lambda_{k>1}$ gradually increase for increasing interactions. At $g_d\gtrsim 0.8$, the $CS$ is reached with $N$ natural orbitals (almost) equally populated [Fig.~\ref{fig:den1d}(b)]. It is a hallmark of the many-body properties of the $CS$: the high-order density matrices, $\rho^{(p)}(p>1)$ are not a product of densities $\rho^{(1)}(x,x')$ [Fig.~\ref{fig:den2d}]. Thus observables such as the correlation functions and the full distribution functions (see below) require a many-body model while classical models fail, see Sec.~S5~\cite{SI}.
 
The maximal value of the crystal order parameter, Eq.~\eqref{OP}, is obtained for the $SF$ with $\Delta=1$, while the $CS$ is identified by the minimum value, $\Delta=\frac{1}{N}=0.125$, hence, characterizing the $CS$ transition.
Fig.~\ref{fig:phase_fdf}(a) shows the value of $\Delta$ as a function of $g_d$ and $V$, which clearly displays the transition from the $SMI$ to the $CS$.

For small $g_d$ and $V$, the bosons are condensed into a $SF$ and $\Delta \approx 1$. With increasing $g_d$ and/or $V$, the $SMI$ forms; fragmentation and -- consequently -- a diminution of $\Delta$ is seen. A further increase of $g_d$ ($>0.8$) decreases $\Delta$ gradually towards its minimum value ($\Delta = 0.125$) for all values of $V$ marking the onset of the $KCS$. When $\Delta$ reaches its minimum, the maximally (eightfold) fragmented $CS$ is reached; the orderings of the $CS$, analyzed in the following, are the $KCS$ or the $DWCS$. Importantly, by analyzing $\Delta$ alone, the $KCS \rightarrow DWCS$ transition cannot be identified.

To identify the $KCS \rightarrow DWCS$, we use the population imbalance of even (e) and odd (o) sites, defined as
\begin{equation}
\Theta = \frac{1}{N} \sum_{e,o} \langle n_o \rangle - \langle n_e \rangle , \label{EO}
\end{equation}
with $\langle n_e \rangle$ and $\langle n_o \rangle$ their respective population.
$\Theta$ is maximal for the density modulation corresponding to the $DWCS$. Fig.~\ref{fig:phase_fdf}(b) shows $\Theta$ as a function of $g_d$ and $V$. For small values of $g_d$, the localization of the atoms in the central wells leads to small values of $\Theta$. For larger $g_d$, $\Theta$ decreases because of the uniform density of the $SMI$. As $g_d$ increases further, $\Theta$ reaches its maximum value for the $KCS\rightarrow DWCS$ transition, which shows a stronger dependence on $V$ compared to $SMI\rightarrow KCS$ transition. A shallower lattice potential favors the $KCS\rightarrow DWCS$ transition at lower $g_d$ while deeper lattices require larger $g_d$, Fig.~\ref{fig:phase_fdf}(b). 
The order parameters $\Delta$ and $\Theta$ quantify  state diagram of any finite-size system as they are applicable for any finite $N$ and $S$.

We now propose a general experimental protocol to detect all emergent states and thereby the state diagram of crystal orders using standard imaging~\cite{Bakr:09,Sherson:10}. These single-shot measurements correspond to a projective measurement of the wavefunction. Ideally, the images contain an instantaneous sample of the position of all $N$ particles distributed according to the $N$-particle probability distribution $\vert \Psi \vert^2$. 
Here, we compute a set of single-shot measurements from our MCTDH-B groundstate wavefunctions~\cite{sakmann16,lode17,tsatsos17,Mistakidis18b} and evaluate the full distribution functions of the position-dependent particle number operator, i.e., we quantify the probability $P_n(x)$ to detect $n$ particles at positions $x$ [Fig.~\ref{fig:phase_fdf}(c--e)].

In the delocalized $SMI$ several particles can be detected in the same site with a significant probability: $P_n(x)$ are nonzero for $n\leq2$. When the $KCS$ is reached at $g_{d}\approx 0.8$, the bosons become completely localized resulting in $P_{n\geq 2}(x)\approx0$. 
The transition from the $SMI$ via the $KCS$ to the $DWCS$ with increasing interaction strength $g_d$ is characterized unequivocally through the analysis of the full distribution functions $P_0(x)$, $P_1(x)$ and $P_2(x)$. While $P_0$ and $P_1$ exhibit the distribution patterns of the $KCS$ and the $DWCS$, $P_2\rightarrow0$ signals the bosons' complete isolation in crystal states -- independent of the crystal ordering. 
The results of Fig.~\ref{fig:phase_fdf}(c--e) show a good agreement with the ones of Fig.~\ref{fig:den1d}(a) and Fig.~\ref{fig:phase_fdf}(a--b), demonstrating the $KCS$ and $DWCS$ transition for the same values of $g_{d}$.
The simultaneous presence of density-wave order in $P_{0/1}$ and isolation, $P_{n \geq 2} \rightarrow 0$, can thus experimentally identify the $KCS$ and the $DWCS$ (See \cite{SI} Sec.~S5, for $P_{n \geq 3}\approx 0$). The FDF as shown in Fig.~\ref{fig:phase_fdf}(c--e) is thus sufficient for detecting all the crystal orders and emergent states and therby the order-parameter and state-diagram. 
Since the full distribution function has been experimentally implemented \cite{betz:11}, our protocol thus provides a general and experimentally feasible way to explore the plethora of crystal orders of dipolar bosons in a lattice. 

We emphasize that our crystal-order detection protocol using full distribution functions makes no reference to $N$ or $S$: it thus represents a viable method for any finite-size system. 

We note that finite-size cold-atom systems necessarily exhibit pinning, thus precluding difficulties arising due to sliding phases that may be seen for periodic boundary conditions (S9 ~\cite{SI}). Moreover, the required accurate single-shot imaging techniques with high detection efficiency and close-to-single-atom detection are available for ultracold atoms \cite{Bakr:09}.

An alternative experimental protocol using the variance of single-shot measurements to quantify the order parameter $\Delta$~\cite{lode17,chatterjee17a} intertwined with a binning of them to quantify the even-odd-imbalance $\Theta$ is described in ~\cite{SI} Sec.~S7.

Beyond the static properties, the exploration of collective excitations like roton modes which have been theoretically predicted~\cite{Maluckov12,Maluckov13,Santos03,Ronen07,Lasinio13} and experimentally observed~\cite{Mottl12,Chomaz18,Petter19} would be of further interest.

We expect low lying phonon-like excitations in our many-body system, similarly to ~\cite{Maluckov13}. In contrast to~\cite{Maluckov13}, the dependence of the crystal spacing on the strength of the dipole-dipole interactions alters the dispersion relation and, therewith, the phonon modes in our setup. 

\acknowledgments{BC acknowledges the financial support from Department of Science and Technology, Government of India under DST Inspire Faculty fellowship. AUJL and CL acknowledge financial support by the Austrian Science Foundation (FWF) under grant No. P-32033-N32, and M-2653, respectively. JS acknowledges funding from the Wiener Wissenschafts- und Technologie Funds (WWTF) project No. MA16-066.
Computation time on the HPC2013 cluster of the IIT Kanpur and the HazelHen and Hawk clusters at the HLRS Stuttgart, as well as  support by the state of Baden-Württemberg through bwHPC and the German Research Foundation (DFG) through grants no INST 40/467-1 FUGG (JUSTUS cluster), INST 39/963-1 FUGG (bwForCluster NEMO), and INST 37/935-1 FUGG ( bwForCluster BinAC) is gratefully acknowledged.}


\begin{thebibliography}{100}
\bibitem{Note_Hubbard} Hubbard models employ a tight-binding approximation suitable in situations where the energy scales  are determined by a lattice potential. For a crystal state, the arrangement of the bosons are determined by dipolar interactions and not by necessarily by the lattice potential. In particular, the wavefunction forms structures on length-scales smaller than the 
lattice constant.
\bibitem{Note_recoil} We make all quantities dimensionless by expressing them in terms of the lattice recoil energy for particles of mass $M$, $E_{R}=\hbar^{2}\kappa^{2}/2M$. This is equivalent to
setting $\hbar=M=\kappa=1$. All quantities become dimensionless with the length given in units of $\kappa^{-1}$.
\bibitem{Note_parameter} $\alpha=0.05$ corresponds to $a_\perp\approx 0.37$ and an aspect ratio $\gamma = \dfrac{a_{\parallel}}{a_\perp} \approx 42.5$. We impose hard-wall boundary conditions at $x=\pm 5\pi/2$  to restrict our lattice to $S=5$ sites.
\end{thebibliography}
\end{document}


\title{Supplementary Information:\\
	Detecting One-Dimensional Dipolar Bosonic Crystal Orders via Full Distribution Functions }
\author{Budhaditya Chatterjee}
\email{bchat@iitk.ac.in}
\affiliation{Department of Physics, Indian Institute of Technology-Kanpur, Kanpur 208016, India}
\author{Camille L\'ev\^eque}
\email{camille.leveque@tuwien.ac.at}
\affiliation{Vienna Center for Quantum Science and Technology, Atominstitut, TU Wien, Stadionallee 2, 1020 Vienna, Austria}
\affiliation{Wolfgang Pauli Institute c/o Faculty of Mathematics, University of Vienna, Oskar-Morgenstern Platz 1, 1090 Vienna, Austria}
\author{J\"org Schmiedmayer}
\email{Schmiedmayer@atomchip.org}
\affiliation{Vienna Center for Quantum Science and Technology, Atominstitut, TU Wien, Stadionallee 2, 1020 Vienna, Austria}
\author{Axel U. J. Lode}
\email{auj.lode@gmail.com}
\affiliation{Institute of Physics, Albert-Ludwig University of Freiburg, Hermann-Herder-Strasse 3, 79104 Freiburg, Germany}
\affiliation{Vienna Center for Quantum Science and Technology, Atominstitut, TU Wien, Stadionallee 2, 1020 Vienna, Austria}

\maketitle

This Supplementary Information discusses the MCTDHB method in Sec.~\ref{Sec:mctdhb}, the one-body momentum density in Sec.~\ref{Sec:rhoK}, the kinetic, potential, and interaction energies in Sec.~\ref{Sec:Energy}, shows results for larger lattice size in Sec.~\ref{Sec:Larger},
illustrates complementary possible crystal orders for different lattice sizes and particle numbers with a classical model in Sec.~\ref{Sec:Classical}, as well as the higher-order distribution functions of the particle number operator in real space, $P_{n>2}(x)$, and the full distribution functions of the particle number operator in momentum space in Sec.~\ref{Sec:FDF}.  An  alternate detection protocol for the phase diagram [Fig.~3(a--b) of the main text] using the variance of single-shot images intertwined with a binning analysis is demonstrated in Sec.~\ref{Sec:SS}. The $1^{st}$ order spatial Glauber correlation function and mean-field results are shown in Sec.~\ref{Sec:MF}. Sec.~\ref{Sec:pbc} shows the results for lattice with periodic boundary condition.

\section{MCTDHB method}\label{Sec:mctdhb}

For our ab-initio real-space quantum many-body simulations, we use the multiconfigurational 
time-dependent Hartree for boson method (MCTDHB) \cite{alon08,streltsov07}, implemented in 
the MCTDHX package \cite{ultracold,axel1,axel2}. This method was recently benchmarked with
experimental observations \cite{tsatsos17} and it was reviewed in Rev.~\cite{lode:19} .

In the MCTDHB method, the many-body wavefunction is expanded as a linear combination of time-dependent permanents, 
\begin{equation}
\vert \Psi \rangle = \sum_{\vec{n}} C_{\vec{n}} (t) \vert \vec{n}; t \rangle. \label{Ansatz}
\end{equation}
The permanents are time-dependent, symmetrized many-boson basis states, constructed via the distribution of $N$ bosons in $M$ distinct orthonormal single-particle orbitals $\lbrace \phi_\alpha(\bm{r};t); \alpha=1,...,M \rbrace$
\begin{equation}
\vert \vec{n} ; t \rangle = \frac{1}{\sqrt{\prod_{\alpha=1}^M n_\alpha !}} \prod_{\alpha=1}^M \left( b_\alpha^\dagger(t) \right)^{n_\alpha} \vert \textrm{vac} \rangle,
\end{equation}
Here, the vector notation is used to denote the bosonic occupation numbers $\vec{n}=(n_1,...,n_M)^T$, fulfilling $N=\sum_\alpha n_\alpha$.  
The number of possible configurations for $N$ bosons in $M$ orbitals is $\binom{N+M-1}{N}$ and 
is equal to the number of (complex-valued) coefficients $C_{\vec{n}}$ in Eq.~\eqref{Ansatz}. 

We now invoke the time-dependent variational principle on the ansatz Eq.~\eqref{Ansatz} to obtain its time-evolution: a set of non-linear integro-differential equations of motion for $M$ orbitals $\lbrace \phi_\alpha(\bm{r};t); \alpha=1,...,M \rbrace$ and linear equations for the $\binom{N+M-1}{N}$ coefficients. Note that the two sets of equations are coupled together.
The simultaneous and self-consistent solution of the equations provides the time-evolution of the coefficients and the orbitals and hence that of the many-body wave function.

The MCTDHB is inherently a time-dependent method. To obtain the ground state, we propagate the MCTDHB equations in imaginary time. 

In our computations, we use $M=8-12$ orbitals to achieve convergence. We used $256$ grid points with FFT DVR representation. Each run is converged at least to the $8^{\text{th}}$ decimal of the total energy value.  

\section{One-body momentum density}\label{Sec:rhoK}

\begin{figure}
	\includegraphics[width=0.5\textwidth]{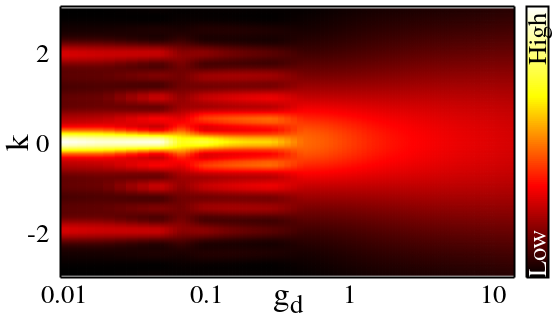}
	\caption{ \textbf{One-particle momentum density $\rho(k)$ as a function of interaction strength $g_d$.} The distinct central peak observed for small interactions demonstrates the spatial delocalization in the $SMI$ state. As $g_d$ increases, the spatial localization corresponding to crystallization renders the momentum distribution delocalized.}
	\label{fig:den1dK}
\end{figure}

Here, we analyze the one-particle momentum density for the same parameters as in Fig.~1 of the main text to assess the spatial coherence of the state, see Fig.~\ref{fig:den1dK}.

For small to moderate interaction strengths ($g_d \lessapprox 0.8$), the density exhibits a central momentum peak surrounded by a delocalized background distribution or small-amplitude peaks in a stripe-like topology. The amplitude and the width of the central peak reduce when the interaction strength $g_d$ increases. The presence of a localized central peak indicates the presence of the delocalized $SF$ fraction that coexists with the localized $MI$ state which in turn is responsible for the delocalized background distribution. When the transition to the crystal state occurs at $g_d \approx 1$, the momentum distribution completely loses the stripe-like topology forming a uniformly delocalized distribution: the crystal state is characterized by strong spatial confinement and decoherence, i.e., a vanishing superfluid fraction.

\section{Energy}\label{Sec:Energy}
\begin{figure}
	\includegraphics[width=0.5\textwidth]{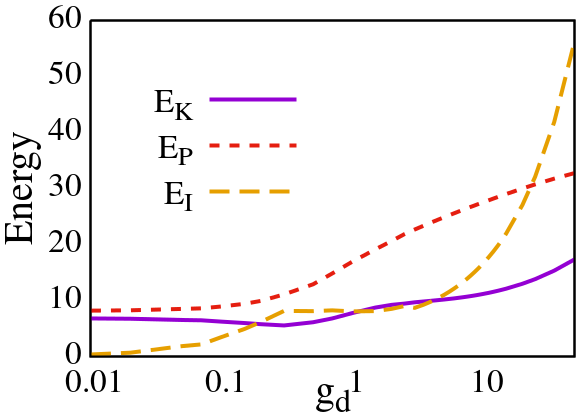}
	\caption{The kinetic, potential, and interaction energies, $E_K$, $E_P$, and $E_I$, respectively, as a function of interaction strength $g_d$ for $N=8$, $V=8$, and $S=5$. While the kinetic and potential energy $ E_K + E_P$ dominate for small to intermediate interactions, the interaction energy $E_I$ clearly dominates for large $g_d$.}
	\label{fig:energy}
\end{figure}

The emergence of distinct quantum phases in dipolar interacting ultracold bosons is a result of the interplay between the interaction, kinetic, and potential energies. In this section we discuss this interplay of energies for the same system as shown in the main text, i.e., $N=8$, $V=8$, $S=5$, as a function of the interaction strength, see Fig.~\ref{fig:energy}. 

The kinetic and potential energy clearly dominate over the interaction energy at small to intermediate interaction strength $g_d\lesssim 0.3$. This dominance is responsible for the formation of the $SF$ and $SMI$ states. The $KCS$ state forms in the region, where the interaction energy and the kinetic energy are comparable, but smaller than the potential energy, i.e., $E_K\approx E_I < E_P$ in Fig.~\ref{fig:energy} for $g_d \in [0.3,2]$.  For larger interaction energies the $KCS$ order becomes energetically unfavorable and the $DWCS$ state emerges. This $DWCS$ state gradually melts as the interactions increase; when the interaction energy $E_I$ eventually dominates over the kinetic and potential energies, $E_I> E_P> E_K$, a ``pure'' crystal state $CS$ is formed. 

\section{Larger lattice size}\label{Sec:Larger}

In the main text, we showed result for $S=5$ and $N=8$. To demonstrate the generality of our results, we here show the calculations for larger lattice size: $S=7$, $N=11$. 

\begin{figure}
	\includegraphics[width=0.5\textwidth]{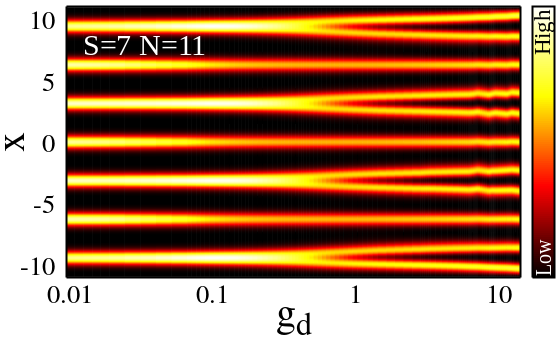}
	\caption{One-body density $\rho(x)$ as a function of the dipolar interaction strength $g_d$ for $S=7,N=11$ at lattice depth $V=8$. The system shows a crystal transition from $SMI$ to a $DWCS$ state. The $KCS$ state is absent.}
	\label{fig:S7}
\end{figure}

Figure \ref{fig:S7} shows  the one-body density $\rho(x)$ for $S=7,N=11$ as a function for the interaction strength $g_d$.
We observe a crystal transition from a $SMI$ state to a $DWCS$ state. The $KCS$ state is absent since the larger lattice size does not have the required kinetic energy to form the $KCS$. Hence we find that for our present incommensurate setup with hard-wall boundaries,  the  $KCS$ is a finite-size feature only, as it fades away for larger systems, in accordance with our classical model. Furthermore, while for our incommensurate setup,  $S=5$ and $N=8$, we do find a $KCS$, we do not observe it on lattices commensurately filled with dipolar bosons \cite{chatterjee17a,chatterjee17b}.

\section{Possible crystal states}\label{Sec:Classical}
In this section, we investigate a complementary model of classical dipolar particles in a lattice in order to highlight that there are several possible ways that dipolar particles may arrange their positions in a lattice as a function of the strength of the dipolar interactions between them. 

The energy of classical dipoles in a lattice is given as
\begin{equation}
E_{\text{class}}(x_1,...,x_N)=\sum_{i=1}^N V_{ol}(x_i) + \sum_{i<j} V_{int}(x_i-x_j).\label{eq:class}
\end{equation}
Here, $x_k$ is the position of the $k$-th particle, $V_{ol}(x)=Vsin^2(\kappa x)$ is the lattice potential energy, and $V_{int}=\frac{g_d}{\vert x_i - x_j\vert^3 + \alpha}$ is the dipolar interaction between particles $i$ and $j$. As in the main text, we use $\alpha=0.05$.

Since we consider the ground state of classical particles (at rest), there's no kinetic energy contribution. Consequently, the kinetic crystal state ($KCS$) observed in the main text for quantum particles, is not present in our classical results below. However, we are able to demonstrate that the density-wave crystal state ($DWCS$) is a generic feature for classical dipolar particles in lattices of different sizes and different filling factors.

\subsection{Classical model verification}
\begin{figure}
	\includegraphics[width=0.45\textwidth]{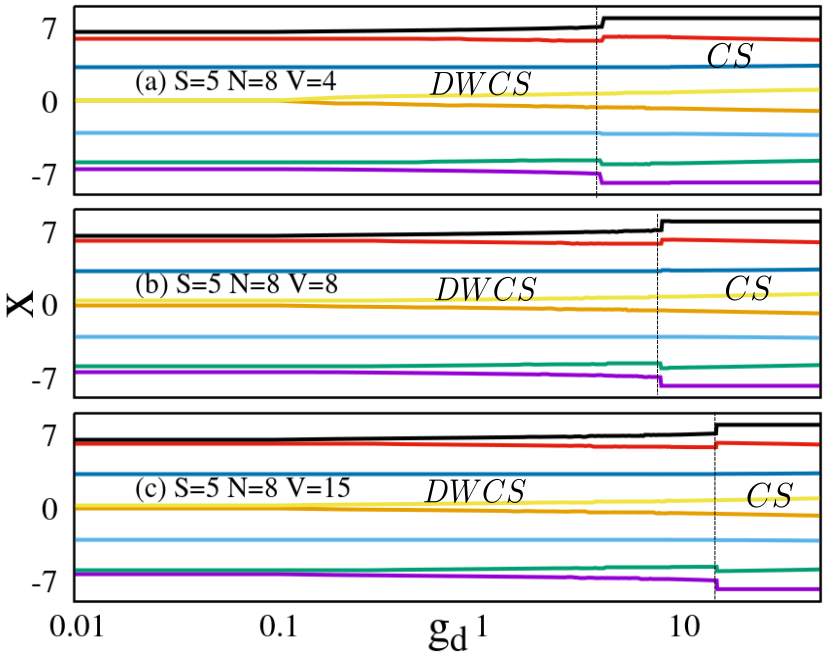}
	\caption{Verification of classical description. Plotted here, the classical position of the atoms $x$ as a function of the interaction strength $g_d$ for different values of lattice depth $V$. The classical description displays different crystal arrangements: the $DWCS$ state and the transition to the $CS$ for strong interactions. Note the absence of the $KCS$ state as well as the $SMI$ state that is seen for the quantum case in the main text Fig.~1.}
	\label{fig:class1}
\end{figure}

We now discuss the results for the positions of classical dipolar particles obtained from minimizing the energy, Eq.~\eqref{eq:class}. To verify the classical model, we begin our investigation with the same configuration, $N=8$ particles in $S=5$ sites, as discussed in the main text for the quantum case. In Fig.~\ref{fig:class1}, we show the positions $x_1,...,x_8$ that minimize $E_{\text{class}}$ as a function of the interaction for three values of the lattice depth, $V=5,8$ and $15$.

The positions of classical dipoles in a lattice, due to the absence of kinetic energy do not feature a kinetic crystal state like the one observed for the quantum particles in the main text. However, the  $DWCS$ is obtained also in the classical model with the same ordering of particles ($[11],1,[11],1,[11]$) similarly to the quantum case in the main text. We infer that we can use our classical model to investigate the possible $DWCS$ orders -- also for other configurations of the number of particles $N$ and the number of lattice sites $S$.

\subsection{Density-wave crystal states for different filling factors}
\begin{figure}
	\includegraphics[width=0.45\textwidth]{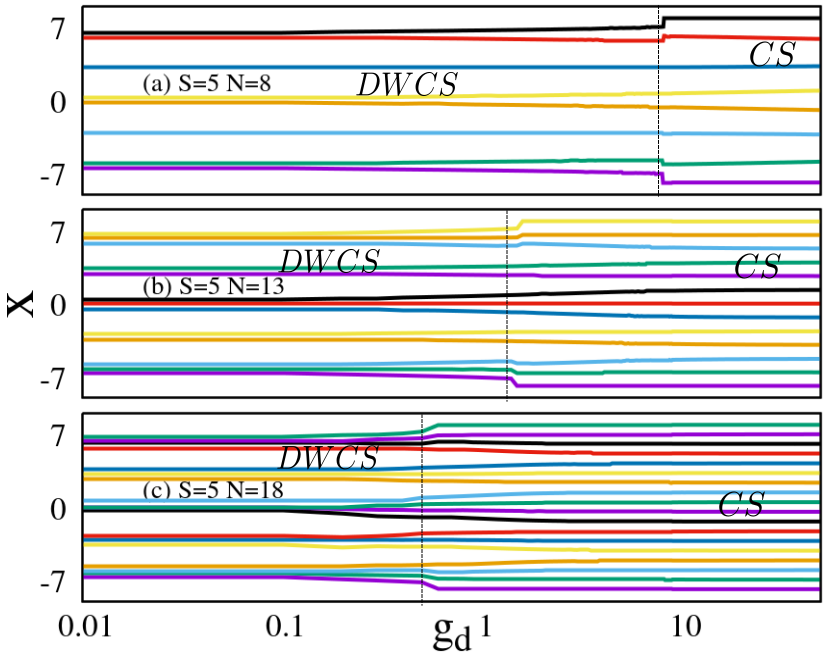}
	\caption{Classical crystal orders for different incommensurate fillings. The (classical) position of the atoms $x$ as a function of the interaction strength $g_d$ is plotted for various atom numbers, $N=8,13,18$, keeping the same $S=5$ lattice size. The different fillings display different $DWCS$ configurations  $[11],[1],[11],[1],[11]$ in (a), $[111],[11],[111],[11],[111]$ in (b), and $[1111],[111],[1111],[111],[1111]$ in (c)] as well as the transition from the $DWCS$ to the $CS$ state.  }
		\label{fig:class2}
\end{figure}

We now discuss the density-wave crystal states that we obtain from our classical model, Eq.~\eqref{eq:class}, for the same number of lattice sites as in the main text, $S=5$, but for different incommensurate filling factors, see Fig.~\ref{fig:class2}.

The $DWCS$ states with orders $[11],[1],[11],[1],[11]$ and $[111],[11],[111],[11],[111]$ and $[1111],[111],[1111],[111],[1111]$ are observed for  $N=8$ and $N=13$ and $N=18$ particles in $S=5$ wells  in Fig.~\ref{fig:class2} panels (a),(b), and (c), respectively. Note that, for a larger particle number, the transition from a $DWCS$ to a $CS$ occurs for lower values of interaction $g_d$. This is because interaction energy per site grows faster compared to the potential energies as a function of particle number at a fixed value of $g_d$.

\subsection{Density-wave crystal states for different lattice sizes}
\begin{figure}
	\includegraphics[width=0.45\textwidth]{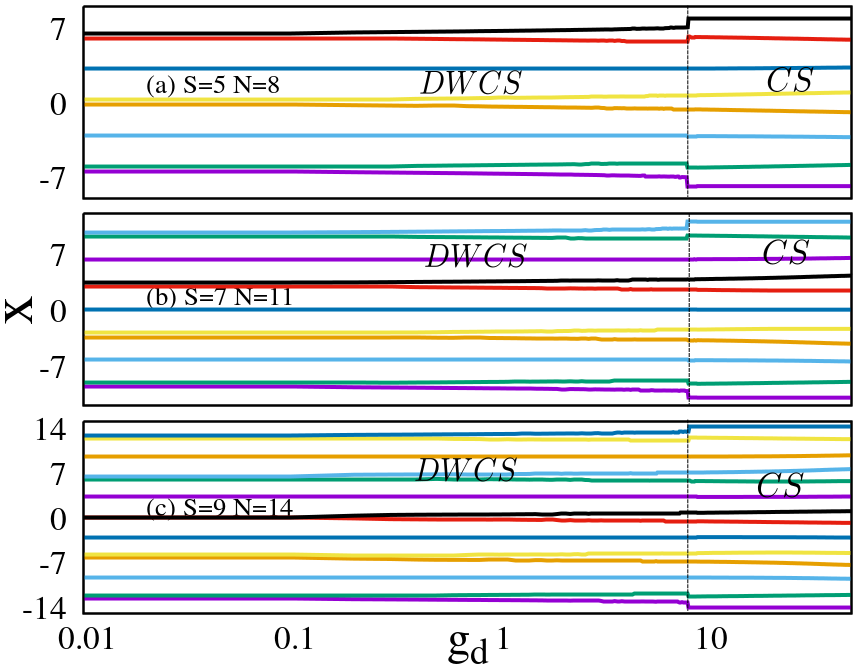}
   \caption{Crystal orders for different lattice sizes. The (classical) position of the atoms $x$ as a function of the interaction strength $g_d$ is plotted for different lattices $S=5,7,9$ keeping the same filling factor.  The $DWCS$ as well as the transition to the $CS$ is seen for all lattice sizes considered.  }
   	\label{fig:class3}
\end{figure}

We now discuss the density-wave crystal states that we obtain from our classical model, Eq.~\eqref{eq:class}, for larger lattice sizes than in the main text ($S=7,9$), but for a similar incommensurate filling factor, see Fig.~\ref{fig:class3}.

It is clearly seen that the $DWCS$ with the configuration $...,[11],1,[11],1,[11],...$ prevails also for the case of lattices with a larger number of sites.

In summary, we find that there is a rich variety of density-wave crystal states of dipolar particles, already in the classical model that we investigated here. This underlines the importance of the experimental protocols that we put forward to detect these states in quantum systems.

\section{Full distribution functions in momentum space  $P_n(k)$ and in real space for larger $P_{n>2}$}\label{Sec:FDF}

\subsection{Full distribution functions in momentum space}
\begin{figure}
 \includegraphics[width=0.5\textwidth]{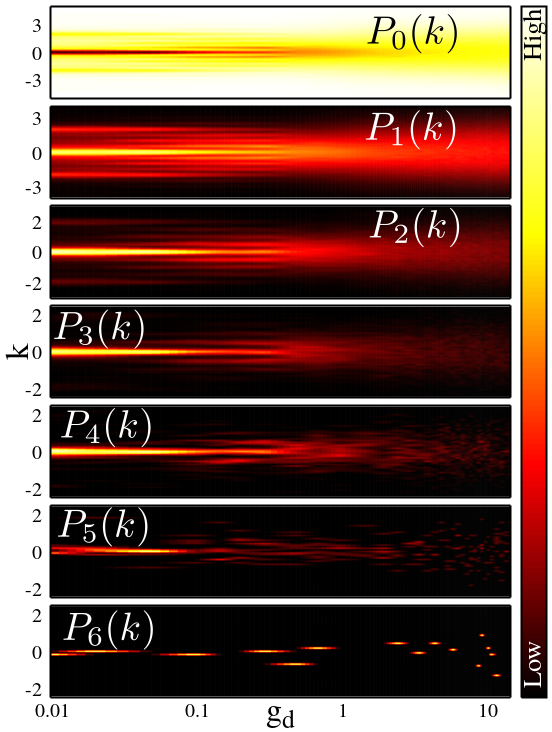}
 \caption{Full distribution functions $P_n(k)$ in momentum space for the same parameters as in Fig.~(3)(c--e) of the main text.}
 \label{fig:SS_FDFk}
\end{figure}

In Fig.~\ref{fig:SS_FDFk}, we show the probability $P_n(k)$ to find $n$ particles with momentum $k$ for the same parameters as in Fig.~(3)(c--e) of the main text.
$P_{0}(k)$ and $P_{1}(k)$ shows identical distribution with the momentum density [Fig. 1(b)] of the main text but with opposite signs. Increasing $n$ shows a distinct narrowing of the distribution. For higher-order distributions $P_{n>3}(k)$, the maximum probability is centered at $k=0$ but falls off sharply and the background is strongly reduced. We remark here, that the total number of events in the plots for $P_{5}(k)$ and, especially, for $P_{n\geq 6}(k)$ is extremely small; thus the probability to find $5$ or more particles at the same momentum  is also extremely small. For $P_{5}(k)$ the maximum count is $\mathcal{O}(10)$ and for $P_{6}(k)$ it is $\mathcal{O}(1)$ out of $N\times N_{shots} = 8\times 10000 = 80000$. 

\subsection{Higher order distribution functions in real space $P_{n>2}(x)$}
\begin{figure}
 \includegraphics[width=0.5\textwidth]{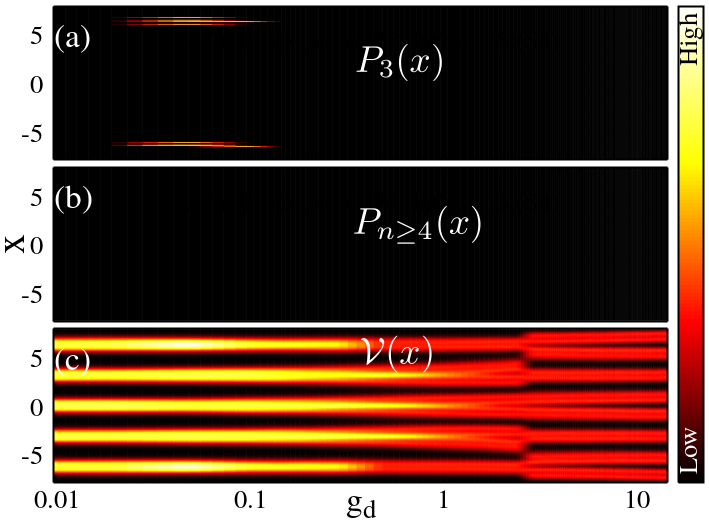}
 \caption{(a--b) Higher order distribution functions in real space $P_{n>2}(x)$ for the same parameters as in Fig.~(3)(c--e) of the main text. (c) Single shot variance as a function of position $\mathcal{V}(x)$ for $V=8$. }
 \label{fig:higherP}
\end{figure}

In Fig.~\ref{fig:higherP}(a--b), we show the probability $P_n(x)$ to find $n$ particles at position $x$ for the same parameters as in Fig.~(3)(c--e) of the main text for $P_{n>2}(x)$. Unlike that of the momentum space, the $x$-space probability becomes zero  for $P_{n>2}(x)$. There is a very small region where $P_{3}(x)\neq0$ while $P_{n\geq4}(x)=0$ for all $x$.  

\section{Detection of the crystal density wave from the variance and imbalance in single-shot images}\label{Sec:SS}

Alternative to the analysis of the full distribution functions shown in the main text, the emergent phases and the phase diagram for the one-dimensional lattices, incommensurately filled with dipolar bosonic atoms, can also be experimentally detected using the single-shot variance in combination with the single-shot expectation of the imbalance parameter $\Theta$ (Eq.~(3) in the main text). 

To characterize the transition from the  $SMI$ state to the $CS$, we use the variance $\mathcal{V}(x)$ of simulated single-shot measurements in position space, i.e., deviations of each single-shot measurement from the mean-value of many single-shot samples [see Refs.~\cite{lode17,chatterjee17a} for the mathematical definition of $\mathcal{V}(x)$]. 

\begin{figure}
	\centering
	\includegraphics[width=0.9\linewidth]{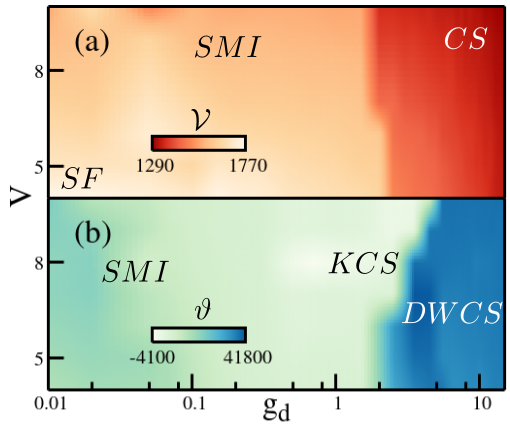}
	\caption {\textbf{Determining the $CS$ and the $DW$ transitions using single-shot simulations}. (a) Variance of single-shots in real-space $\mathcal{V}(g_d,V)$ as a function of $g_d$ and $V$. Every data point corresponds to variance $\mathcal{V}$ computed from $10000$ single-shot samples. $\mathcal{V}$ is maximum for $SF$ decreasing for $SMI$ and reaches a minimum for $CS$. The distinct transition to minimum $\mathcal{V}$ clearly displays the $CS$ transition. (b) Imbalance obtained from binning single-shot simulation $\vartheta$.  The distinct transition to high $\vartheta$ demonstrates the experimental detection of the $DW$ transition.}
	\label{fig:SS} 
\end{figure}

The variance strongly depends on the degree of localization, i.e., on how much the positions of the bosons fluctuate from image to image~\cite{chatterjee17a}. For increasing localization, the variance decreases. Since the $CS$ corresponds to a maximal localization of the system, $\mathcal{V}(x)$ attains its minimum for the $CS$. 
Fig.~\ref{fig:higherP}(c) displays the variance $\mathcal{V}(x)$ as a function of position $x$ for $V=8$. A clear reduction of the variance $\mathcal{V}(x)$ is seen with interaction $g_d$ as the system transition from the $SMI$ to the $CS$. The variance integrated over space, $\mathcal{V}=\int{\mathcal{V}(x)\text{d}x}$, hence can be used to determine the $CS$ transition experimentally.
Fig.~\ref{fig:SS}(a) displays the integrated variance $\mathcal{V}$ as a function of $g_d$ and $V$. 

The $SF$ state is maximally delocalized, thus having a maximum value of $\mathcal{V}$. With increasing interactions, the  $SMI$ state is reached and the partial localization of the bosons at the lattice sites leads to a decrease of $\mathcal{V}$. When the $CS$ is reached at large values of $g_{d}$, the bosons are completely localized. The vanishing overlap between the bosons forces $\mathcal{V}$ towards its minimum in the $CS$.
The results of Fig.~\ref{fig:SS}(a) are in good agreement with the ones of Fig.~3(a) of the main text, predicting the transition to the $CS$ state for the similar values of $g_{d}$ and $V$. The variance of single-shot images can thus be used to experimentally determine accurately the transition to the crystal state.

In order to detect the transition to the $CDW$ state, we evaluate the population imbalance parameter $\Theta$, Eq.~(3) of the main text, by binning the single-shot images: we count for each single-shot the number of atoms detected in the vicinity of each minimum of the lattice. From the binned single shot densities, the single shot imbalance $\vartheta$ is calculated. When $\vartheta$ is obtained from the average of many single-shot realizations, converges to the imbalance parameter $\Theta$.

The value of $\vartheta$ as a function of $g_d$ and $V$ is shown in Fig.~\ref{fig:SS}(b). The  transition to the $DW$ state is clearly visible in the value of $\vartheta$: the state diagram of the even-odd imbalance parameter $\Theta$ [Fig.~3(b) of the main text] closely resembles the behavior of the imbalance parameter $\vartheta$ obtained from the binning of single-shot measurements [Fig.~\ref{fig:SS}(b)]. Our simulations of single-shot measurements correspond to the experimental imaging process. We have thus demonstrated a viable experimental protocol to detect both, the crystal state and the crystal density wave transitions and, therefore, a protocol to determine the state diagram of lattices that are incommensurately filled with dipolar bosons.

\section{Correlation function and Mean field results}\label{Sec:MF}

\subsection{$1^{st}$ order spatial Glauber correlation functions}

In the main text, we analyzed the $2^{nd}$ order spatial  correlation functions in Fig. 2(e--h). In this section, we display the corresponding $1^{st}$ order spatial  correlation function.

The $1^{st}$ order correlation function $g^{(1)}(x_1,x_1^{\prime}) = \frac{ \rho^{(1)}(x_1,x_1^{\prime})} {\sqrt{\rho(x_1)\rho(x_1^{\prime})}}$ quantifies the $1^{st}$ order coherence of the system; $|g^{(1)}|=1$ implies perfect coherence and  $|g^{(1)}|=0$ no coherence. $g^{(1)}(x_1,x_1^{\prime})$ and thus also provides  an assessment of the deviation of the many-body state from a completely condensed (mean-field) state since the condensed state is completely coherent.

\begin{figure}
	\includegraphics[width=1.0\linewidth]{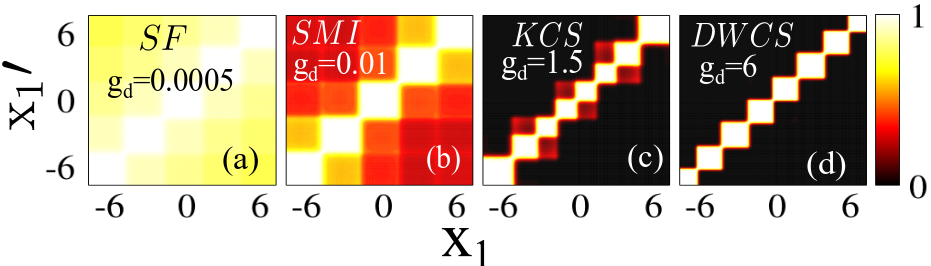}
	\caption{$1^{st}$ order correlation function $g^{(1)}(x_1,x_1^{\prime})$ (a) $g_d = 0.0005$, $SF$ shows (almost) complete coherence with $|g^{(1)}|\approx1$ throughout. (b) $g_d = 0.01$, $SMI$ the coherent regions are diagonal. (c) $g_d = 1.5$, $KCS$ coherent blocks splits due to crystallization. (d) $g_d = 6$, $DWCS$ completely split coherent blocks centered on each boson position.  }
	\label{fig:g1x}
\end{figure}

In $SF$ state [$g_d = 0.0005$, Fig.~\ref{fig:g1x}(a)], the system, being largely delocalized, shows substantial coherence, i.e., predominantly uniform $|g^{(1)}|\approx1$ for all $x_1,x'_1$. For the $SMI$ state, $g_d = 0.01$, the partial localization in the lattice produces the coherence blocks seen in panel (b) of Fig.~\ref{fig:g1x}. For stronger interactions $g_d =1.5$ [$KCS$, panel (c)], the coherence blocks split due to the crystallization of the state with a remaining coherence between the bosons in doubly occupied sites. For very strong interactions $g_d = 6$ in the $DWCS$, the coherence blocks are completely split and  are centered at each boson's position, Fig.~\ref{fig:g1x}(d).

\subsection{Mean-field results}

This section displays the mean-field (MF) results for the two-body densities and the Glauber correlation functions and compares them with the many-body (MB) results of the main text.

\begin{figure}
	\includegraphics[width=1.0\linewidth]{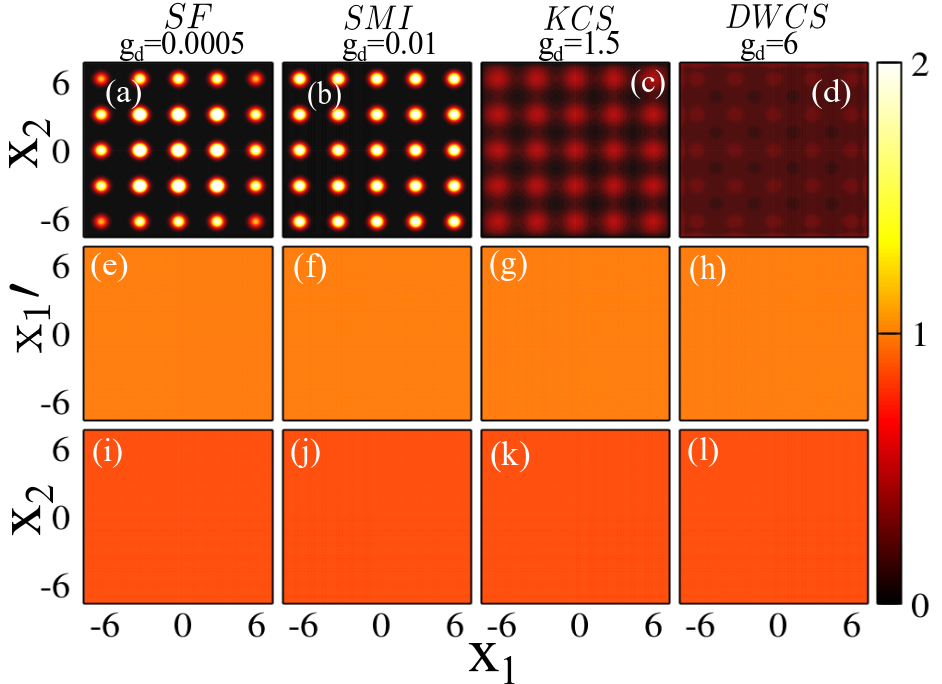}
	\caption{Mean field results for  two-body densities and the correlation functions. (a)--(d) The two-body densities $\hat{\rho}_2(x_1,x_2)$.
	(e)--(h): The $1^{st}$ order spatial correlation function $|g^{(1)} (x_1,x_1')|$ and (i)--(l): the $2^{nd}$ order correlation function (diagonal) $|g^{(2)} (x_1,x_2)|$.}
	\label{fig:MF_den2d}
\end{figure}

At $g_d = 0.0005$, the system is $SF$ and significantly condensed and the mean-field (MF) calculation gives the correct two-body densities $\rho^{(2)}(x_1,x_2)$, Fig.~\ref{fig:MF_den2d}(a). For the $SMI$ state ($g_d = 0.01$), the mean-field results deviates from the MB results [compare Fig.~\ref{fig:MF_den2d}(b) to Fig.~2(b) of the main text] and does not show a depletion of the diagonal. For stronger interactions $g_d \geq 1.5$, the MF results completely deviates from the MB results [compare Fig.~\ref{fig:MF_den2d}(c),(d) to Fig.~2(c),(d) of the main text].

For the $1^{st}$ [$g^{(1)}$] and $2^{nd}$ [$g^{(2)}$] order Glauber correlation functions, the MF results, by construction, show complete coherence and thus are only valid for extremely weak interactions $g_d\approx 0$, see Fig.~\ref{fig:MF_den2d}(e)--(l).  
 Already for $g_d = 0.0005$, the MF results diverges from the MB ones, and for stronger interactions, the MF results completely deviates from the MB results for both $|g^{(1)}|$ and  $|g^{(2)}|$ [compare Fig.~\ref{fig:MF_den2d}(i)--(l) to Fig.~2(e)--(h) of the main text and Fig.~\ref{fig:MF_den2d}(e)--(h) to Fig.~\ref{fig:g1x}(a)--(d), respectively].
Thus we can conclude that the results in the main text arise inherently from quantum many-body effects and cannot be obtained from a mean-field analysis. 

\section{Periodic Boundary condition}\label{Sec:pbc}

The main text discussed the lattice system with hard-wall boundary conditions. Here, we show the one-body density for a lattice with periodic boundary condition (PBC). 

\begin{figure}
	\includegraphics[width=0.5\textwidth]{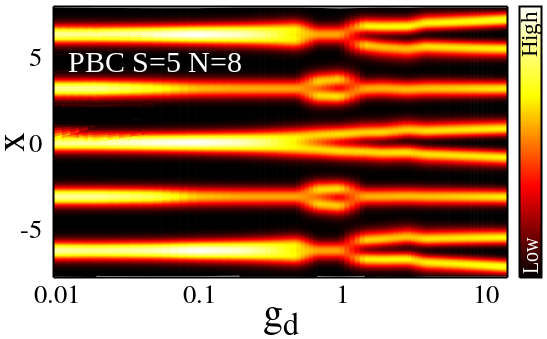}
	\caption{One-body density $\rho(x)$ as a function of the dipolar interaction strength $g_d$ for $S=5,N=8$ for a lattice with \textit{periodic boundary condition} (PBC) at lattice depth $V=8$. }
	\label{fig:S5pbc}
\end{figure}

For a lattice with periodic boundary condition, all the states - the $SMI$, the $KCS$, and $DWCS $ states are present. However, instead of the usual $DWCS$ state, we obtain an interesting sliding $DWCS$ state; the axis of symmetry of the CDW phase changes or "slides", as all possible positions of the density structure with respect to the central site at $x=0$ are energetically degenerate. This sliding is a hallmark of the setup with periodic boundary condition.

In Fig.~\ref{fig:S5pbc}, the obtained densities have been shifted such that the symmetry with respect to $x=0$ was restored.

For $g_d\sim1$, the $KCS$ state is obtained for a small range of interactions. 

Subsequently, for $g_d>9$, we obtain the $DWCS$ state. In summary, we find that all crystal orderings that we found with hard walls in the main text do prevail for the case of periodic boundary conditions for $S=5, N=8$, and $V=8$.